\documentclass[%
 reprint,
bibnotes,
 amsmath,amssymb,
prb,
]{revtex4-2}
\usepackage[colorlinks=true, urlcolor=blue, pdffitwindow=true, linkcolor=blue, citecolor=blue, pdfstartview={FitH}]{hyperref}
\usepackage{graphicx}
\usepackage{dcolumn}
\usepackage{bm}

\usepackage{epstopdf}
\usepackage[sort&compress]{natbib}
\begin{document}


\title{Intersubband optical absorption in the GaN/AlN qunatum wire with donor-doping}

\author{Sami Ortakaya}
 \altaffiliation[Also at ]{Shipito Address:\\
             444 Alaska Avenue Suite $\#$BKF475, Torrance 90503, CA, USA}
 \email{sami.ortakaya@yahoo.com}
\affiliation{Institute for Globally Distributed Open Research and Education (IGDORE), Gothenburg, Sweden
}%
\date{{\color{black}\today}}

\begin{abstract}
In this study, we investigate the optical absorption within the conduction-subbands of a cylindrical GaN/AlN quantum wire. We analyze the optical absorption rate and the real part of the dielectric function for both quantum wire (QWR) and quantum dot (QD) structures in the presence of donor-impurity states. The results cover that the density of states associated with free motion in the QWR structure leads to a lower optical absorption compared to the QDs, as evidenced by the optical spectra. This study advances the understanding of GaN/AlN heterostructures by offering a comprehensive analysis of the optical properties of QWRs and QDs, especially under donor-doping situation. As a result, the outcomes provide a clear evidence of the effect of semiconductor nanostructures on the optical properties of optoelectronic devices.
\end{abstract}

\maketitle
\section{Introduction}
III-Nitrides (i.e, GaN, AlN, InN and their alloys) have attracted significant interest due to their exceptional optical and electronic properties, making them establishing them as key components in optoelectronics. Especially, GaN-based quantum nanostructures offer unique advantages, such as high exciton binding energy, strong quantum confinement effects, and enhanced optical absorption in the ultraviolet (UV) region \cite{zhang2023}. These characteristics make them suitable for applications in light-emitting diodes \cite{kobayashi}, laser diodes \cite{wang2024}, and high-power photodetectors \cite{quach2022}.  

In recent years, considerable research has been devoted to investigating the electronic and optical properties of GaN-based heterostructures \cite{samihocaefendi, gan1, gan2} . The addition of an AlN shell around a GaN core significantly modifies the potential landscape, leading to a deep confinement potential in the conduction subband. This deep confinement can strongly influence the optical absorption and dielectric response of the system. Several theoretical and experimental studies have examined the impact of core/shell geometries on the electronic structure and interband transitions in GaN/AlN structure \cite{sami,gs1,gs2,gs3}. However, the precise effects of core size variations on optical absorption and the real part of the dielectric function remain an active area of investigation.  

The quantum states of charge carriers in hetero-nanostructures have been extensively studied, yielding significant results on  their electronic and optical properties \cite{mommadi, nguyen, kasapoglu, zeiri, namin}. Numerous investigations have also focused on the energy levels and binding energies in QWRs. Baser {\it et al} \cite{baraz2011} have investigated the effects of hydrostatic pressure on the electronic properties of InGaN/GaN cylindrical QWR. In the realm of optical properties, studies have successfully derived optical gain spectra for InAsP/InP QWRs \cite{peter2016}. Similarly, research has been conducted to calculate the variations in energy levels based on the impurity-position within GaN cylindrical QWR \cite{duque2014}. While these studies have laid the groundwork for modeling such systems, they often lack an updated framework for accurately describing optical properties. In this context, there remains a need for further exploration, particularly in calculating the dielectric function spectrum and comparing the optical properties of QWR and QD.

In this study, we analyze the optical absorption properties and the real part of the dielectric function for GaN/AlN QWR with cylindrical GaN core layers of 3 nm and 4 nm radius. The addition of an AlN shell induces significant modifications in the electronic structure, leading to enhanced quantum confinement effects. By employing theoretical modeling and numerical calculations, we aim to provide a deeper understanding of how these structural variations influence the optical response of GaN/AlN quantum wires. The results of this work could contribute to the optimization of GaN-based heterostructures for future optoelectronic applications.  

\section{Theoretical Model}
\subsection{GaN/AlN Quantum Wire}
\begin{figure}[!hbt]\centering
\scalebox{.35}{\includegraphics{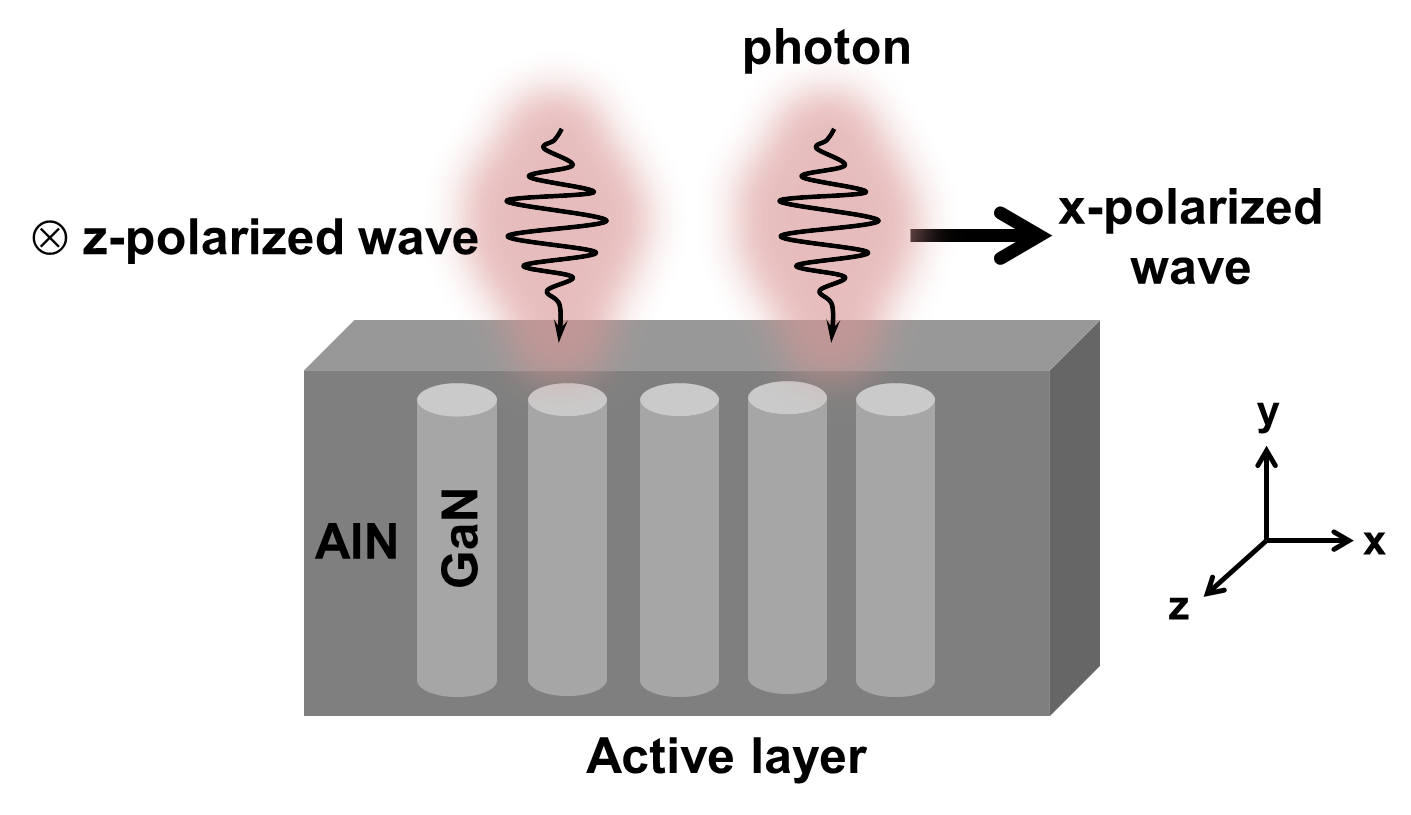}}
\scalebox{.35}{\includegraphics{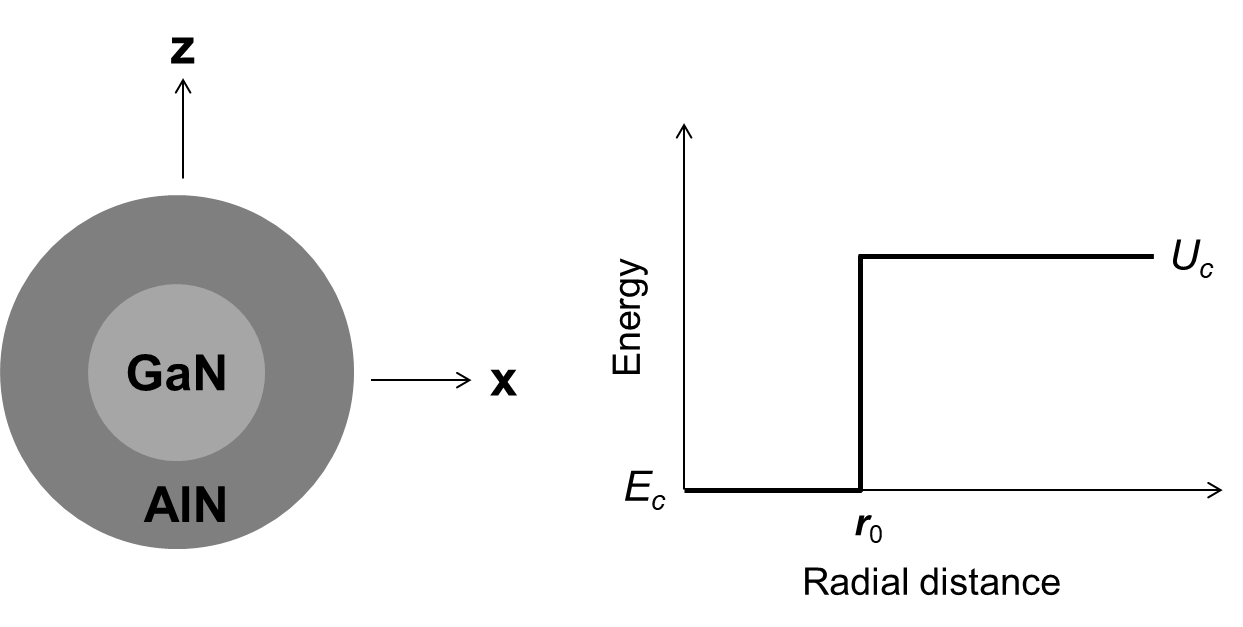}}
\caption{Typical GaN/AlN quantum wire based layers and conduction band profile.}\label{wire1}
\end{figure}

The GaN core and AlN shell are designed as the active layer of an optical absorptive device, and the optical system is illustrated in Figure \ref{wire1}. GaN QWRs allow electronic transport by polarization of the optical wave. This layer is a "pure active layer", and in the absence of n- and p-contacts, the conduction band profile is obtained without simulating the potential (i.e., in the absence of Schrödinger-Poisson coupling). 

The GaN/AlN core/shell structure and the conduction band profile are also shown in Figure \ref{wire1}. In this structure, while the optical wave polarized along the direction of free carrier motion is not absorbed, the polarized-wave in other directions can be easily absorbed. We consider the isolated cylindrical zinc-blende GaN QWR of radius $r_{0}$ surrounded by large band-gap material, AlN. The coordinate origin is taken at the quantum wire axis and the $y$-axis is defined to be the QWR axis. The donor doping is a hydrogen model which is located in the wire axis. The electron bound states and the corresponding wave functions are obtained by Schrodinger equation under effective-mass aproximation.

Within framework of the effective-mass approximation, for
a hydrogen donor which is located at center of QWR, the electronic Hamiltonian in cylindrical coordinates $(r,\,\theta,\,y)$ can be written as

\begin{equation}\label{hat}
\hat{H}=-\frac{\hbar^2}{2}\vec{\nabla}\cdot\left(\frac{1}{m_e^* (r)}\vec{\nabla}\right)+U_{c}-\frac{e^2}{4\pi\epsilon\left|{\bm{r}}\right|},
\end{equation}
where $\hbar$, $m_e^*$ and $e$ represent Planck's constant,  electron effective-mass and elementary charge, respectively. $U_c$ denotes the confinement potential energy obtained from dicontinuties for rate \%76 in the GaN/AlN heterostructure including GaN core of dielectric permittivity $\epsilon=\epsilon_{0}\epsilon_r$. Considering that a band gap difference between GaN and AlN, we can write confinement potential, 
\begin{equation}
U_c=
\begin{cases}
    0,& r<r_0\\
    0.76\times[\Delta E_g],             & r\geq r_0.
\end{cases}
\end{equation}
Note that the $\epsilon_r$ is dielectric constant of bulk material which is different from static dielectric constant of considered model. For a donor located at the wire axis covers axis $y$- in the free motion, we take $|\bm{r}|=\sqrt{r^2+y^2}$. The wavefunctions related to eigenvalues $E_{n\ell}$ of the 2D-polar Hamiltonian without Coulomb interaction, can be given by
\begin{eqnarray}\label{psps}
&&\Psi_{n\ell}(r,\,\theta)=e^{i \ell \theta}R(r),\\
&&(\ell=0,\,\pm1,\,\pm2,\,\dots\;\;\text{and}\;\; n=1,\,2,\,3,\,\dots)\nonumber 
\end{eqnarray}
where $e^{i \ell \theta}$ and $R(r)$ are angular and radial parts of wave function is given in the following unnormalized form,
\begin{equation}\label{rr}
R(r)=
\begin{cases}
    J_{\ell}(\kappa_w r),& r<r_0\\
    K_{\ell}(\kappa_b r),             & r\geq r_0,
\end{cases}
\end{equation}
where 
\begin{equation}
\kappa_w^2 = \frac{2m_{e}^* E_{n\ell}}{\hbar^2},\qquad \kappa_b^2 =  \frac{2m_{e}^* (U_c-E_{n\ell})}{\hbar^2}.
\end{equation}
The radial function $R(r)$ is defined by $\ell$th Bessel function $J_{\ell}$ and modified Bessel function $K_{\ell}$. In order to obtain energy eigenvalues of donor-doped QWR, we define following trial wavefunction,
\begin{equation}
\Psi_{t}(r,\,\theta,\, y)=N \Psi_{n\ell}(r,\,\theta)e^{-a(r^2+y^2)},
\end{equation}
where $a$ and $N$ denote variational parameter and normalization constant, respectively. The wavefunction $\Psi_{n\ell}(r,\,\theta)$ is 2D quantized eigenstate which is defined in Equation (\ref{psps}).  The energy level $\overline{E}_{n\ell}$ of the QWR with donor-impurity can be calculated by minimizing process in the form
\begin{equation}
\overline{E}_{n\ell}=\min_{a}\frac{\langle \Psi_t |\hat{H}|\Psi_t \rangle}{\langle \Psi_t |\Psi_t \rangle},
\end{equation}
 where $\hat{H}$ is Hamiltonian operator is defined in Equation (\ref{hat}). The energy values and corresponding wave functions derived via variational method are pioneer quantities in obtaining optical coefficients, thus allowing for the examination of  optical properties related to the considered quantum system.
 
 \subsection{Solutions of the Liouville equation for dipole interaction}
In order to get the mean dipole moment for optical coefficients, we consider that the incident optical wave includes electric field, $\mathcal{E}_{0}$ which interact with electron in the subband. The Hamiltonian related to interaction, can be given by
\begin{eqnarray}
H_{\rm int}=-M\mathcal{E}_{0}=\left(\begin{array}{ccccc} 
M_{11}\mathcal{E}_{0} & M_{12}\mathcal{E}_{0} & \cdots & M_{1 f}\mathcal{E}_{0} & \cdots\\ 
M_{21}\mathcal{E}_{0} & M_{22}\mathcal{E}_{0} & \cdots & M_{2 f}\mathcal{E}_{0} & \cdots\\
\cdots\ & \cdots\ & \cdots\ & \cdots\ & \cdots\\
M_{ f1}\mathcal{E}_{0} & M_{ f2}\mathcal{E}_{0} & \cdots & M_{ ff}\mathcal{E}_{0} & \cdots\\
\cdots\ & \cdots\ & \cdots\ & \cdots\ & \cdots
 \end{array}\right),\nonumber\\
 {}
\end{eqnarray}
where we define dipole matrix element in the form: $M_{if}=\left<\Psi_{i}|er|\Psi_{f}\right>$ with pure states $\Psi_{i}$ and $\Psi_{f}$. Mean dipole moment in unit volume $V$ represents the polarization through a monochromatic wave of frequency, $\omega$.  Namely, we write the polarization of the form $P(\omega)=\frac{\left<M\right>}{V}.$ On the basis of density matrix related to the mean dipole moment, we have
\begin{equation}
\left<M\right>=\sum_{i,\,f}\left<f|\rho|i\right> \left<i|M|f\right>={\rm Tr}(\rho M),
\end{equation}
where $N\times N$ density matrix is given by
\begin{eqnarray}
\rho=\left(\begin{array}{ccccc} 
\rho_{11} & \rho_{12} & \cdots & \rho_{1 f} & \cdots\\ 
\rho_{21} & \rho_{22} & \cdots & \rho_{2 f} & \cdots\\
\cdots\ & \cdots\ & \cdots\ & \cdots\ & \cdots\\
\rho_{f1} & \rho_{f2} & \cdots & \rho_{f f} & \cdots\\
\cdots\ & \cdots\ & \cdots\ & \cdots\ & \cdots
 \end{array}\right).
\end{eqnarray}
The allowed levels including unperturbed Hamiltonian $H_0$ and interaction within time-$t$ evolution, can be launched the following Liouville's equation \cite{Shen2003}
\begin{equation}
\frac{\partial\rho(\omega,\,t)}{\partial t}=-\frac{{\rm i}}{\hbar}\left[H_{0} - M\mathcal{E}_{0},\,\rho\right]-\Gamma\rho-\rho\Gamma,
\label{liouville}
\end{equation}
where $\Gamma$ is 'damping term for scattering \& phononic context. Putting the well-known matrix trace operation, we should write \cite{optik1}
{\setlength\arraycolsep{0.1em}\begin{eqnarray}
P(\omega,\,t)&=&\frac{1}{V}{\rm Tr}(\rho M)\nonumber\\&=&\frac{1}{V}\big[\rho_{11}M_{11}+\rho_{12}M_{21}+
\rho_{21}M_{12}+\rho_{22}M_{22}\big].\nonumber\\
\label{trace}\end{eqnarray}}
The time evolution which is represented by Equation (\ref{liouville}), is re-written through matrix element $\rho_ {21}$. This element is obtained via following form
\begin{widetext}
{\setlength\arraycolsep{0.1em}\begin{eqnarray}
\frac{\partial \rho_{21}}{\partial t}&=&\Big<2\Big|\frac{\partial \rho}{\partial t}\Big|1\Big>\nonumber\\
&=&\frac{-{\rm i}}{\hbar}
\Big[\left<2|(H_{0}-M\mathcal{E}_0)\rho|1\right>-\left<2|\rho(H_{0}-M\mathcal{E}_0)|1\right>\Big]-\Big[\left<2|\Gamma\rho|1\right>+\left<2|\rho\Gamma|1\right>\Big]\nonumber\\
&=&\frac{-{\rm i}}{\hbar}
\Big[\big<2\big|H_{0}(|1\big>\big<1|+|2\big>\big<2|)\rho\big|1\big>-\big<2\big|M\mathcal{E}_{0}(|1\big>\big<1|+|2\big>\big<2|)\rho\big|1\big>\Big]
\nonumber\\
&+&\frac{{\rm i}}{\hbar}
\Big[\big<2\big|\rho(|1\big>\big<1|+|2\big>\big<2|)H_{0}\big|1\big>-\big<2\big|\rho(|1\big>\big<1|+|2\big>\big<2|)M\mathcal{E}_{0}\big|1\big>\Big]\nonumber\\
&-&\Big[\left<2|\Gamma(|1\big>\big<1|+|2\big>\big<2|)\rho|1\right>+\left<2|\rho(|1\big>\big<1|+|2\big>\big<2|)\Gamma|1\right>\Big],
\end{eqnarray}}
\end{widetext}
where eigenvalues $E_1$ ve $E_2$ should provide that $\left<1|H_0|1\right>$=$E_ {1}$, $\left<2|H_0|2\right>$=$E_ {2}$. Furthermore, non-diagonal elements are given as $\left<2|H_0|1\right>$=0, $\left<1|H_0|2\right>$=0. Then, we should have
{\setlength\arraycolsep{0.1em}\begin{eqnarray}
\frac{\partial \rho_{21}}{\partial t}&=&\frac{-{\rm i}}{\hbar}
\Big[(E_{21}+(\Delta M)\mathcal{E}_{0})\rho_{21}+(\rho_{22}-\rho_{11})M_{21}\mathcal{E}_{0}\Big]\nonumber\\&-&(\Gamma_{22}+\Gamma_{11})\rho_{21}-\Gamma_{21}(\rho_{11}+\rho_{22}).
\label{ilke}\end{eqnarray}}
Here,  $E_{21}=E_2 - E_1$ and $\Delta M=M_{11}-M_{22}$. The phenomenological operator for scattering mechanisms shows relaxation time with non-diagonal and diagonal elements for $\Gamma_{21}$=$\Gamma_{12}$=0 and $\Gamma_{11}=\Gamma_{22}=\gamma/2$, respectively. Here, well-known approach to non-diagonal elements of unperturbed Hamiltonian, is to consider agreement with non-diagonal elements of the damping, $\Gamma$ . So that, Equation (\ref{ilke}) reads
{\setlength\arraycolsep{0.1em}\begin{eqnarray}\label{roo21}
\frac{\partial \rho_{21}}{\partial t}&=&\frac{-{\rm i}}{\hbar}
\Big[(E_{21}+(\Delta M)\mathcal{E}_{0}-{\rm i}\hbar\gamma)\rho_{21}\nonumber\\&+&(\rho_{22}-\rho_{11})M_{21}\mathcal{E}_{0}\Big].
\end{eqnarray}}
Putting the monochromatic optical wave, we have electric field in the following form
\begin{equation}
\mathcal{E}=\mathcal{E}_{0}\cos\omega t = \mathcal{E}_{0}\exp( \mathrm{i}\omega t)+\mathcal{E}_{0}\exp(-\mathrm{i}\omega t),
\end{equation} 
we also consider that the input frequency equals to output, then we have resonance polarization,
\begin{equation}P_{\rm res}(\omega,\, t)=\left(\chi^{(1)}\mathcal{E}_{0}+
\frac{3}{4}\chi^{(3)}\mathcal{E}_{0}^{3}+\cdots\right){\rm e}^{{\rm- i}\omega t}.\end{equation}
We also take that the stationary solution is
\begin{equation}
\rho(\omega,\,t)=\rho(\omega){\rm e}^{{\rm- i}\omega t}.\label{cozgen1}
\end{equation}
Inserting the other elements, we obtain that
\begin{subequations}
{\setlength\arraycolsep{0.1em}\begin{eqnarray}
\rho_{21}({\omega})&=&\frac{(\rho_{11}-\rho_{22})M_{21}\mathcal{E}_{0}}{E_{21}-\hbar\omega+(\Delta M)\mathcal{E}_{0}-{\rm i}\hbar\gamma},\\
\rho_{12}({\omega})&=&\frac{(\rho_{11}-\rho_{22})M_{12}\mathcal{E}_{0}}{E_{21}+\hbar\omega+(\Delta M)\mathcal{E}_{0}+{\rm i}\hbar\gamma},\\
\rho_{11}(\omega)&=&
\frac{(M_{21}\rho_{12}-M_{12}\rho_{21})\mathcal{E}_{0}}{\hbar\omega+{\rm i}\hbar\gamma},\\
\rho_{22}(\omega)&=&
\frac{(M_{12}\rho_{21}-M_{21}\rho_{12})\mathcal{E}_{0}}{\hbar\omega+{\rm i}\hbar\gamma}.
\end{eqnarray}}
\end{subequations}
So that, the polarization of considered subband transition leads to
{\setlength\arraycolsep{0.1em}\begin{eqnarray}
P_{\rm res}(\omega)&=&\frac{(\rho_{11}-\rho_{22})|M_{12}|^{2}}{V}\bigg[\mathcal{E}_{0}\left(\frac{1}{E_{+}}+\frac{1}{E_{-}}\right)\nonumber\\&+&(\Delta M)\mathcal{E}_{0}^{2}\left(\frac{1}{\hbar\omega+{\rm i}\hbar\gamma}\right)\left(\frac{1}{E_{+}}-\frac{1}{E_{-}}\right)\bigg],\nonumber\\
\label{anadenklem1}
\end{eqnarray}} 
where, $E_{-}=E_{21}-\hbar\omega+(\Delta M)\mathcal{E}_{0}-{\rm i}\hbar\gamma$ and $E_{+}=E_{21}+\hbar\omega+(\Delta M)\mathcal{E}_{0}+{\rm i}\hbar\gamma$. The optical wave of frequency $\omega$ causes to the polarization function,
{\setlength\arraycolsep{0.1em}\begin{eqnarray}
P_{\rm res}(\omega,\,t)&=&\epsilon_{0}\Big[A_{1}\chi_{\rm res}^{(1)}(\omega)\mathcal{E}_{0}+A_{3}\chi_{\rm res}^{(3)}(\omega)\mathcal{E}_{0}^ 3\nonumber\\&+&A_{5}\chi_{\rm res}^{(5)}(\omega)\mathcal{E}_{0}^5 +\cdots\Big]{\rm e}^{{\rm- i}\omega t},
\end{eqnarray}}
where $A_{1}$=1, $A_{3}$=$\frac{3}{4}$ and $A_{5}$=$\frac{5}{8}$ are obtained from trigonometric relations.

Considering that Equation (\ref{anadenklem1}) has the stationary solution,
$$P_{\rm res}(\omega,\,t)\to\epsilon_{0}\chi_{\rm res}^{(1)}(\omega)\mathcal{E}_{0}{\rm e}^{{\rm- i}\omega t}$$
and the occupancy of the subbands can be given by
$$\rho_{11}-\rho_{22}\to\rho_{11}^{(0)}-\rho_{22}^{(0)}=(N_{\rm el}+1)-(N_{\rm el})=1$$
where $N_{\rm el}$ is particle number. In actual way, we get that $N_{\rm el}=1$ denotes full-occupied ground- and empty excited-states. Then, we get
\begin{equation}
\chi_{\rm res}^{(1)}(\omega)=\frac{|M_{12}|^2}{\epsilon_{0}V}\left[\frac{1}{E_{+}}+\frac{1}{E_{-}}+\frac{(\Delta M)\mathcal{E}_{0}}{\hbar\omega+{\rm i}\hbar\gamma}\left(\frac{1}{E_{+}}-\frac{1}{E_{-}}\right)\right].
\end{equation}
In order to get an optical transition between quantum states instead of the classical dipole resonance, we re-write resonance form in the dielectric constant as follows:
\begin{equation}
\tilde{\epsilon}_{r}=1+\chi^{(1)}+\tilde{\chi}^{(1)}_{\rm res}.
\end{equation}
Within the linear term, the dielectric constant becomes a complex function which is given by
{\setlength\arraycolsep{0.1em}\begin{eqnarray}
\tilde{\epsilon}_{r}&=&1+\chi^{(1)}+\tilde{\chi}^{(1)}_{\rm res}\nonumber\\
&=&1+\chi^{(1)}+\frac{|M_{12}|^2}{\epsilon_{0}V}\Big[\frac{1}{E_{+}}+\frac{1}{E_{-}}\nonumber\\&+&\frac{(\Delta M)\mathcal{E}_{0}}{\hbar\omega+{\rm i}\hbar\gamma}\left(\frac{1}{E_{+}}-\frac{1}{E_{-}}\right)\Big].\label{dielr}
\end{eqnarray}}

\section{Numerical Results}
The energy levels and corresponding wave functions in the subbands have been obtained as variational solutions to the Schrödinger equation for both ground state (10) and excited state (11) where QWR quantum states are represented by $(n\ell)$ For the spectral calculations at 300 K, the material parameters and Fermi-Dirac distribution were used, with optical broadening $\hbar \gamma = 10$ meV. The parameters for GaN and AlN are provided in Table \ref{rtablo}. 

\begin{table}[!hbt]\centering\caption{The recommended values of material parameters for zb-GaN and zb-AlN. Here, $m_0$ is the free electron mass.}\vspace*{10pt}
\begin{tabular}{cccccc}
\hline
{Material parameters} &&&&GaN&AlN\\
\hline
$E_{g}\;(\rm eV)$&&&&$3.2\,^{\rm a}$&$5.3\,^{\rm b}$\\
$m^{*}_{\rm e}/m_{0}$&&&&$0.13\,^{\rm c}$&$0.19\,^{\rm d}$\\
$\epsilon_{\rm st}$&&&&$9.56\,^{\rm e}$&$8.35\,^{\rm e}$\\
$\epsilon_{\infty}$ &&&&$5.3\,^{\rm f}$\\
\hline
\end{tabular}\\
\vspace*{9pt}
\footnotesize$^{\rm a}$ Ref. \cite{degheidy2011}, $^{\rm b}$ Ref. \cite{thompson2001}, $^{\rm c}$ Ref. \cite{fan1996},$^{\rm d}$ Ref. \cite{pugh1999}, $^{\rm e}$ Ref. \cite{persson2007}, $^{\rm f}$ Ref. \cite{boles2017}
\label{rtablo}\end{table}

\begin{figure}[!hbt]\centering
\scalebox{.85}{\includegraphics{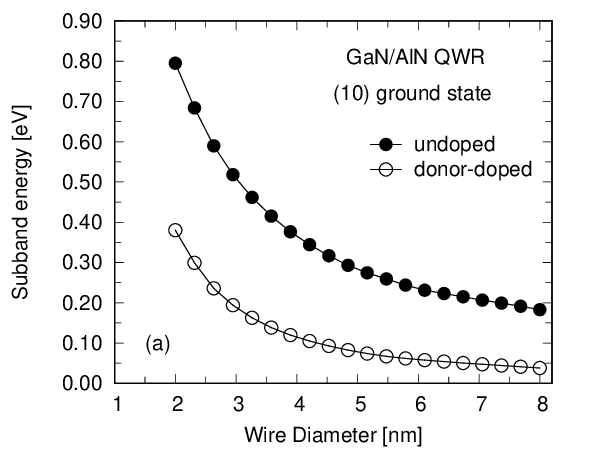}}\\
\scalebox{.85}{\includegraphics{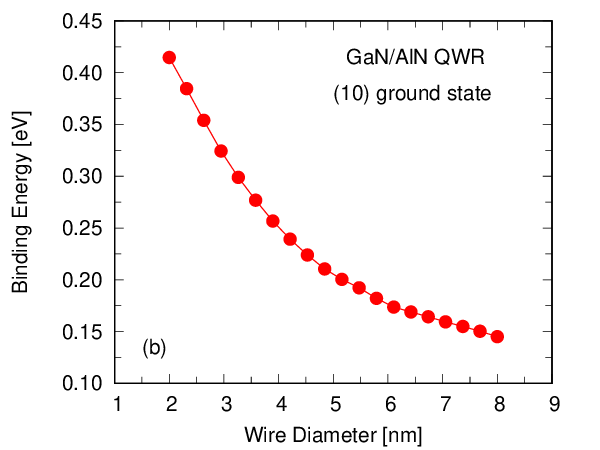}}
\caption{Conduction subband energy levels and binding energy as a function of wire diameter. Here, (10) represents the quantum state for $n=1$ and $\ell=0$.}\label{Eb}
\end{figure}

In the GaN/AlN QWR, the subband energy levels of the ground state in the conduction band, as a function of the cylindrical wire radius, are shown in Figure \ref{Eb}a. As expected, the obtained values of the energy levels decrease with increasing radius for both undoped and donor-doped cases. These variations follow a familiar $\frac{1}{r^2}$ radial behavior. Consequently, a similar trend is observed in the binding energy, as depicted in Figure \ref{Eb}b. The impurity mechanism considered at the central axis, leads to a reduction in energy levels due to Coulomb interaction. As the radius increases, lower binding energies are obtained, indicating that the impurity-ionization process may become more likely as the size of the QWR increases.

\begin{figure}[!hbt]\centering
\scalebox{.85}{\includegraphics{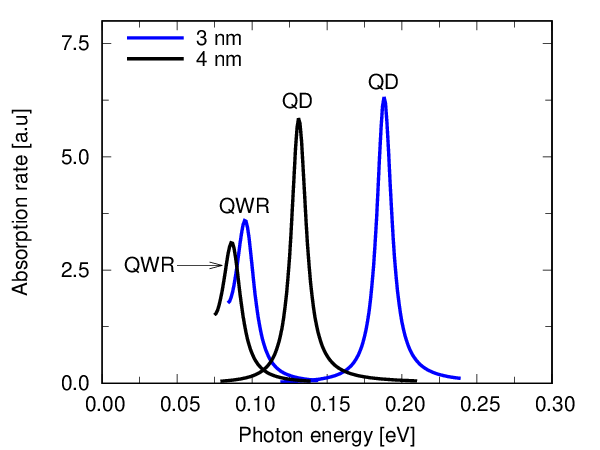}}
\caption{The absorption spectra of $(10)\to(11)$ subband-transition in the GaN/AlN heterostructures. Here (10) and (11) represents $(n\ell)$ quantum states.}\label{abs}
\end{figure}

The absorption rate spectra are plotted in Figure \ref{abs}, comparing quantum dots (QDs) and quantum wires (QWRs). The QD is modeled for spherical GaN/AlN core/shell structures, as adapted from Ref. \cite{sami}. Both systems feature radius values of 3 nm and 4 nm. These values are valid for the core radius of a spherical QD and a cylindrical QWR in a core/shell structure.  The intersubband transitions in the conduction subbands lead to the transition rate is given in following form\cite{fox}
\begin{equation}\label{wif}
W_{i\to f}=\frac{2\,\pi}{\hbar}\,| \langle i|f\rangle |^2 \cdot \delta(E_f-E_i-\hbar\, \omega)\,,
\end{equation}
where $\omega$ is the incident optical frequency in the monochromatic form, and $| \langle i|f\rangle |$ denotes the dipole matrix element between the two energy level. The transition rate can be given in the another way through absorption coefficient, which provides an QD optical transition rate at the ``resonance'' peaks, so it is given by
\begin{equation}\label{wif2}
w[{\rm QD}]=W'_{i\to f}\propto \hbar\omega| \langle i|e\cdot {\bm r}|f\rangle |^2 \cdot \frac{1}{(E_f-E_i-\hbar\omega)^2+(\hbar\gamma)^2},
\end{equation}
where $\gamma$ is the damping term given in Equation (\ref{roo21}). Due to the QWR structure has a partial continuum, we consider the QWR with transition rate,
\begin{equation}
w[{\rm QWR}]=D(E,\,E_i)\times W'_{i\to f},
\end{equation}
where $D(E,\, E_i)$ is density of states given by
\begin{equation}D(E,\, E_i)=\frac{\sqrt{m^*}}{\pi\hbar\sqrt{E-E_i}},\end{equation}
where $E_i$ denotes $i$th subband energy level. For a radius of 3 nm, the energy spectrum exhibits a red shift as the system transitions from lower to higher dimensions.  This is due to the fact that the energy difference between the corresponding levels for the transition energies is greater for a 3 nm radius compared to that of a 4 nm radius. We apply light wave in $x$- or $z$- polarization as given in Figure \ref{wire1}, so we obtain that
\begin{equation}
| \langle i|e\cdot{\bm r}|f\rangle |^2=| \langle i|e\cdot{\bm x}|f\rangle |^2 , \qquad (x=r\cos\theta),
\end{equation}
and we can also choose $z=r\sin\theta$ through the matrix element, $\langle i|e\cdot{\bm z}|f\rangle$. The spectra which is provided by selection rule $\Delta\ell=\pm 1$, obviously reveal distinct differences: the QWR exhibits fewer peaks compared to the QD, and the 3D confinement of QD displays a narrower spectral range. Additionally, a red shift is observed in the transition energies for the QWR relative to the QD, indicating that the use of 2D confinement results in a spectral shift toward lower values in the infrared wavelength range.  This broadening of the spectra in QWR is primarily attributed to the density of states, $D(E,\,E_i)$. In contrast, the QD exhibits larger overlap values due to its 3D confinement space, which enhances the absorption characteristics. 

\begin{figure}[!hbt]\centering
\scalebox{.85}{\includegraphics{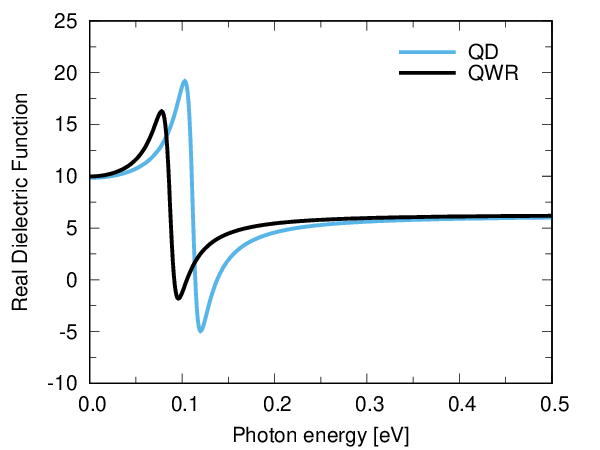}}
\caption{$(10)\to(11)$ intersubband transition related to dielectric function of GaN/AlN QWR and spherical QD adapted from Ref. \cite{sami}. }\label{diel}
\end{figure}

As seen in Figure \ref{diel}, the spectrum representing the real part of the dielectric function yields static dielectric constant values at zero frequency of 10.05 for the GaN/AlN QWR and 9.89 for the GaN/AlN QD structure, both with a radius of 4 nm. The dielectric function is provided in Equation (\ref{dielr}) as a result of the time evolution of the density matrix.
The main reason for the lower values observed in the quantum wire is the 2D confinement in QWRs, as opposed to the 3D confinement in QDs. Due to the 3D confinement in QDs, higher overlap between allowed energy levels is achieved, resulting in spectrum with higher values for QDs. Additionally, the sheet density of the QWR is lower compared to that of the QD. The sheet density is given by 
{\setlength\arraycolsep{0.1em}\begin{eqnarray}
\frac{1}{V}&=&\frac{1}{L_w}\Big[\int D(E,\,E_f)f(E,\,E_f)dE\nonumber\\&-&\int D(E,\,E_i)f(E,\,E_i)dE\Big],\quad L_w=\rm 4.0 \times 4.0\, \rm nm^2\nonumber\\
\end{eqnarray}}
where $f(E,\,E_i)$ and $D(E,\,E_i)$ represent Fermi-Dirac distribution and density of states, respectively. Note that we take a typical value, 10 meV corresponding energy difference between Fermi-level and full occupied ground state.
\section{Conclusion}

In this study, the ground state binding energy, optical transition rates, and the real part of the dielectric function have been calculated for a GaN/AlN cylindrical quantum wire. The binding energies exhibited a \(1/r^2\) dependence on the wire radius, while the optical spectrum was compared with that of a quantum dot. The dielectric function spectrum was plotted via revised iterative density matrix. It has been observed that the optical absorption for the GaN-core quantum wire was lower compared to that of the quantum dot, and a similar trend was noted for the real dielectric constant. The GaN/AlN structure, with its deep potential well, allows for the modeling of infrared absorption in a 1D architecture using wide-bandgap materials.

One of the most pioneer aspects of this work is its comparability with previous quantum dot calculations. While the quantum wire yields values close to those of the quantum dot, its numerical values for optical absorption are lower. These results serve as a stepping stone toward achieving optimal data in optoelectronic device design. Future studies utilizing these comparative optical data will provide a crucial methodology for determining device performance. 
\bibliography{refs}

\end{document}